\documentclass[conference]{IEEEtran}
\UseRawInputEncoding
\IEEEoverridecommandlockouts
\usepackage{cite}
\usepackage{tabularx}
\usepackage{textcase}
\usepackage{amsmath,amssymb,amsfonts}
\usepackage{algorithmic}
\usepackage{graphicx}
\usepackage{textcomp}
\usepackage{xcolor}
\usepackage{booktabs, tabularx}
\usepackage{rotating}
\usepackage{comment}
\usepackage{placeins}
\usepackage{hyperref}
\usepackage{textcomp}

\usepackage{stfloats}
\def\BibTeX{{\rm B\kern-.05em{\sc i\kern-.025em b}\kern-.08em
		T\kern-.1667em\lower.7ex\hbox{E}\kern-.125emX}}

\usepackage{array}
\newcolumntype{M}[1]{>{\centering\arraybackslash}m{#1}}
\newcolumntype{N}{@{}m{0pt}@{}}
\usepackage{makecell}

\begin{document}
	
	\title{A Survey of 5G-Based Positioning for Industry 4.0: State of the Art and Enhanced Techniques \\
		%\thanks{This Project has received funding from the European Union’s Horizon 2020 research and innovation programme under the Marie Sklodowska-Curie grant agreement No 956670.}
	}
	
	\makeatletter
    \newcommand{\linebreakand}{%
    \end{@IEEEauthorhalign}
    \hfill\mbox{}\par
    \mbox{}\hfill\begin{@IEEEauthorhalign}
}
\makeatother
	
	\author{
    \IEEEauthorblockN{
        Karthik Muthineni\IEEEauthorrefmark{1}\IEEEauthorrefmark{2}, Alexander Artemenko\IEEEauthorrefmark{1}, Josep Vidal\IEEEauthorrefmark{2}, Montse Nájar\IEEEauthorrefmark{2}
    }
    \IEEEauthorblockA{\IEEEauthorrefmark{1} Corporate Sector Research and Advance Engineering, Robert Bosch GmbH, Renningen, Germany}
    \IEEEauthorblockA{\IEEEauthorrefmark{2} Department of Signal Theory and Communications, Universitat Politècnica de Catalunya (UPC), Barcelona, Spain}
    \IEEEauthorblockA{Email: \IEEEauthorrefmark{1}$\{${karthik.muthineni, alexander.artemenko}$\}$@de.bosch.com, \IEEEauthorrefmark{2}$\{${josep.vidal, montse.najar$\}$@upc.edu}}
}
	
	%\author{\IEEEauthorblockN{Karthik Muthineni\IEEEauthorrefmark{1},
    %Alexander Artemenko\IEEEauthorrefmark{2}, Josep Vidal Manzano\IEEEauthorrefmark{3} and
    %Montse Nájar\IEEEauthorrefmark{4}}
    %\IEEEauthorblockA{Corporate Sector Research and Advance Engineering,
    %Whichever University\\
    %Wherever\\
    %Email: \IEEEauthorrefmark{1}karthik.muthineni@de.bosch.com,
    %\IEEEauthorrefmark{2}alexander.artemenko@de.bosch.com,
    %\IEEEauthorrefmark{3}josep.vidal@upc.edu,
    %\IEEEauthorrefmark{4}montse.najar@upc.edu}}
	
	%\author{\IEEEauthorblockN{Karthik Muthineni}
	%	\IEEEauthorblockA{\textit{Corporate Sector Research and Advance Engineering} \\
	%		\textit{Robert Bosch GmbH}\\
	%		Renningen, Germany \\
	%		karthik.muthineni@de.bosch.com}
	%	\and
	%	\IEEEauthorblockN{Alexander Artemenko}
	%	\IEEEauthorblockA{\textit{Corporate Sector Research and Advance Engineering} \\
	%		\textit{Robert Bosch GmbH}\\
	%		Renningen, Germany \\
	%		alexander.artemenko@de.bosch.com}
	%	\linebreakand
	%	\IEEEauthorblockN{Josep Vidal Manzano}
	%	\IEEEauthorblockA{\textit{Department of Signal Theory and Communications} \\
	%		\textit{Universitat Politècnica de Catalunya (UPC)}\\
	%		Barcelona, Spain \\
	%		josep.vidal@upc.edu}
	%	\and
	%	\IEEEauthorblockN{Montse Nájar}
	%	\IEEEauthorblockA{\textit{Department of Signal Theory and Communications} \\
	%		\textit{Universitat Politècnica de Catalunya (UPC)}\\
	%		Barcelona, Spain \\
	%		montse.najar@upc.edu}
		
	%}
	
	\maketitle
	
	\begin{abstract}
		
		%The next generation of mobile communication 6G is going to enable sensing of the environment as an in-built feature applying the same electro-magnetic waves used for the data communication. This introduces a very strong disruption of the conventional sensing products. With this talk, we want to introduce the new feature of 6G called Integrated Sensing and Communication (ISAC) showing its potential in context of SLAM. Furthermore, we present the state of the art in ISAC as well as its roadmap and first thoughts on related Bosch products. The fifth geenration (5G) mobile communication systems have the potential to estimate the positions, orientations, clock bias of the User Equipment (UE), and map the propagation environments around UE without any additional hardware.
		The fifth generation (5G) mobile communication technology integrates communication, positioning, and mapping functionalities as an in-built feature. This has drawn significant attention from industries owing to the capability of replacing the traditional wireless technologies used in industries with 5G infrastructure that can be used for both connectivity and positioning. To this end, we identify the Automated Guided Vehicle (AGV) as a primary use case to benefit from the 5G functionalities. Given that there have been various works focusing on 5G positioning, it is necessary to analyze the existing works about their applicability with AGVs in industrial environments and provide insights to future research. In this paper, we present state of the art in 5G-based positioning, with a focus on key features, such as Millimeter Wave (mmWave) system, Massive Multiple Input Multiple Output (MIMO), Ultra-Dense Network (UDN), Device-to-Device (D2D) communication, and Reconfigurable Intelligent Surface (RIS). Moreover, we present the shortcomings in the current state of the art. Additionally, we propose enhanced techniques that can complement the accuracy of 5G-based positioning in controlled industrial environments. 
		%The fifth geenration (5G) mobile communication systems have the potential to estimate the positions, orientations, clock bias of the User Equipment (UE), and map the propagation environments around UE without any additional hardware. This document is a model and instructions for \LaTeX. This and the IEEEtran.cls file define the components of your paper [title, text, heads, etc.]. *CRITICAL: Do Not Use Symbols, Special Characters, Footnotes, or Math in Paper Title or Abstract. The 5G technology alone cannot satisfy the accuracy requirements of AGVs in the industrial environments. Therefore, it is necessary to work on new possibilities of improving the positioning accuracy by complementing the 5G technology with non-radio technologies. In this section, we present three improvement techniques that can enhance the 5G-based positioning accuracy, while remaining in congruence with the 5G system. Figure~\ref{figure1} shows the positioning architecture of AGV.
		
	\end{abstract}
	
	
	\section{Introduction}
	The fifth generation (5G) mobile communication systems have the potential to estimate the positions, orientations, and clock bias of the User Equipment (UE) and map the propagation environment around UE without any additional hardware \cite{BAQUERO2022}. This is made possible by taking the advantage of 5G positioning capabilities: high frequencies, large bandwidths, multiple antennas, Ultra-Dense Network (UDN), and Device-to-Device (D2D) communication. Of many application scenarios, where 5G is considered beneficial, industrial automation benefits highly from the 5G positioning capabilities because positioning information can help track the process flow and automate manual tasks. 
	
	The transport vehicles in industries are used to assist the workers in loading/unloading the goods and transporting them to the desired destination within the industry. However, these vehicles are either manually operated or are pre-programmed to move along a specified path, by following the magnetic tape installed on the floor. The vision of Industry 4.0 (I4.0) is to introduce the autonomous feature in transport vehicles, to navigate in manufacturing units and warehouses without manual intervention \cite{DERYCK2020}. Such a transport vehicle is called an Automated Guided Vehicle (AGV), which offers several benefits for industries such as decreased labor costs, improved product safety, and improved work efficiency. These strong points are motivating the industries to deploy AGVs in their facilities to improve the daily operations like transportation of goods in manufacturing units and sorting/picking of items from warehouses. For instance, an autonomous transport system called ActiveShuttle is implemented to transport goods in industrial sites, replacing traditional conveyor belts \cite{BOSCH}. 
	
	The deployment of a large number of AGVs in industries can be seen as a big step in modernizing the industrial transportation systems that pave the way for building future I4.0. However, there are certain problems associated with the deployment of AGVs, that do not allow to use them everywhere. The first problem that AGVs need to overcome is positioning. They need a precise positioning system to help them navigate to the right pickup station to load the goods and also make use of the same positioning system to navigate to the docking station to undock the goods. Table \ref{tab1} shows the requirements of AGV in the industrial environment. The second problem is related to the robustness of wireless communication. AGV communicate with other AGVs and localization servers using wireless communication. Today, most AGVs use WiFi as a communication channel and if the industry has poor wireless connectivity, it affects the entire fleet of AGVs. The third problem is associated with the mapping of plant information. It is necessary to create the map, identifying the positions of pickup/drop stations, charging docks, and parking slots, which AGVs will use during positioning. The process of creating maps itself is a tedious job and it requires experts in the field to conduct several on-site procedures, which might be time-consuming.
	
	To achieve maximum efficiency from AGV, authors believe that industries need strong advances in two technological fields: 1) wireless communication and 2) indoor positioning. To this end, 5G is considered a key enabler for I4.0 due to its capabilities that extend beyond smart devices. In the literature, there are various research works that focus on achieving precise positioning with 5G using different approaches. However, there is a lack in the analysis of their applicability to AGV in the industrial environment. 
	
	The main contributions of this paper are as follows:
	\begin{enumerate}
		\item A comprehensive review of recent developments in 5G-based positioning and presents their shortcomings. 
		%\item This paper summarizes the problems present in the state-of-the localization methods.
		\item Enhanced techniques that have the potential to improve the accuracy of 5G-based positioning for the use case of AGV in a controlled industrial environment.
	\end{enumerate}
	
	\begin{table}[t]
	\caption{Automated guided vehicle (AGV) requirements.}
	\centering
	\begin{tabular}{p{0.15\linewidth}p{0.15\linewidth}p{0.15\linewidth}p{0.15\linewidth}p{0.15\linewidth}}
		\hline
		Positioning  & Orientation & Latency & Obstacle detection & Reference\\
		\hline
		$<$1 cm & 1$^\circ$ & $<$10 ms & Yes & \cite{BOSCH}\\
		\hline
	\end{tabular}
	\label{tab1}
\end{table}

\section{Cramer–Rao Lower Bound as Positioning Performance Indicator}
Indoor positioning can be considered as an estimation problem, where the AGV estimates its position in the industry based on the signals received from multiple base stations. However, the presence of heavy metallic objects in the industry can cause signal reflections and diffractions. Therefore, the estimator function should be capable of estimating the AGV position from noisy observations. The two main criteria for an efficient estimator are: the estimates should be unbiased and should have low variance (from the true value of the parameter). The estimator is said to be unbiased if the mean of its estimates equals the true value of the parameter, which can be verified mathematically. But, verifying the estimator for the low variance criteria is difficult. Cramer–Rao Lower Bound (CRLB) provides the lowest bound or the minimum variance that can be achieved by any unbiased estimator, trying to estimate a particular parameter \cite{JAGAN2004}. 
%The estimator whose variance is close to the CRLB is considered an efficient estimator and is said to meet the low variance criteria. 
The Position Error Bound (PEB) and Orientation Error Bound (OEB) are the two known performance indicators for positioning technology which can be derived by performing the square root of CRLB of position and orientation. 

\section{state of the art}
The requirement laid to network operators by United States Federal Communications Commission (FCC) to provide emergency positioning (E-911) in the mid-1990s and by European Union (EU) Coordination Group on Access to Location Information for Emergency Services (CGALIES) to provide location information (E-112) in 2000 had begun the demand for Location-Based Services (LBS) in cellular networks \cite{DEL2018}. Subsequently, the Third Generation Partnership Project (3GPP) has introduced specifications for positioning methods in cellular networks (2G, 3G, 4G) by designing new approaches to provide position estimates with existing signal structures. %The advantages of using cellular-based positioning are that they do not require the deployment of additional wireless infrastructure and they can achieve both indoor and outdoor positioning simultaneously. 
However, until now, the cellular networks (2G, 3G, and 4G) fell short in providing the positioning accuracy needed by AGV for navigating autonomously in industries, due to noisy observations resulting from multipath fading, Non-Line-of-Sight (NLoS) situations, and due to synchronization error among base stations \cite{DEL2018}. On the other hand, the 3GPP Service and System Aspects (SA) group study on the 5G positioning for indoor environments has targeted the accuracy of decimeters. To this end, in 3GPP Release 16, the Radio Access Network (RAN) group introduced new measurements for positioning, which include the DownLink Time Difference of Arrival (DL-TDoA), UpLink Time Difference of Arrival (UL-TDoA), DownLink Angle of Departure (DL-AoD), UpLink Angle of Arrival (UL-AoA), and Multi-Cell Round Trip Time (RTT). Continuing the target of achieving precise positioning in indoors, Release 17 has introduced methods to enhance the accuracy of measurements and mitigate the timing delays between the base station and UE \cite{3GPPlocation}. 
%The recent research works in 5G positioning show that sub-meter positioning accuracy can be achieved by exploiting the information in MultiPath Component (MPC), based on a new positioning approach called multipath-assisted localization \cite{WITRISAL2016}. 
%Moreover, the 5G network offers licensed, unlicensed, and shared spectrum, providing industries with diverse options for deploying their private networks. The unlicensed spectrum provides large bandwidth, which is useful to achieve accurate Time of Arrival (ToA) measurements. However, uncontrolled interference in the unlicensed spectrum from other systems can show a negative result on positioning accuracy. A controlled industrial environment can overcome this problem. Authors define a controlled environment as limiting the number of devices operating in the same frequency band as that of a 5G network.
In this section, we present the recent developments in 5G-based positioning. Table \ref{tab2} summarizes the recent developments. 
The works presented in the literature are mainly based on simulation results, not all assumptions are realistic, and do not consider heavy multipath conditions in an industrial environment.
	
	%\subsection{Positioning with 5G}
	%The mobile networks standardization body 3GPP has 
	%mentioned about positioning in 5G networks in its draft Release 16, which introduces new reference signals and protocols that are necessary to exchange location information between UE/5G NR and Location Management Function (LMF). Release 16, focuses on enhancing the positing accuracy of multi cell Round Trip Time (RTT), Angle of Arrival (AoA), and Angle of Departure (AoD) measurements.
	%The 5G network offers licensed, unlicensed, and shared spectrum, providing industries with diverse options of deploying their private networks. The unlicensed spectrum provides large bandwidth, which is useful to achieve accurate ToA/TDoA measurements. However, the uncontrolled interference in unlicensed spectrum from other systems can show negative result on positioning accuracy. The controlled industrial environments can overcome this problem. Authors define controlled environments as limiting the number of devices operating in the same frequency band as that of 5G network. In this section, we present the recent developments of 5G-based positioning.
	
	\subsection{5G Millimeter Wave System}
	As per the 3GPP, frequencies in the Frequency Range (FR) 2, starting from 30 GHz are called Millimeter Wave (mmWave) frequencies \cite{3GPPlocation}. At these frequencies, the large available bandwidth provides high time resolution in delay estimate, because of which accurate Time of Arrival (ToA) measurement for MultiPath Component (MPC) can be achieved. However, higher propagation losses at mmWave frequencies are a major challenge. Therefore, it is important to understand the limits of these frequencies for positioning. In \cite{ABUSHABAN2018}, authors derived the CRLB for position and orientation errors. Moreover, authors investigated how the low-scattering nature of mmWave frequencies can deduce the problem of multipath estimation to multiple single-path parameter estimation. In their model, authors considered a single base station with multiple antennas and a UE operating at a carrier frequency of 38 GHz, orientated by an angle of 10$^\circ$. The Fisher Information Matrix (FIM) is computed for the Angle of Arrival (AoA), Angle of Departure (AoD), and ToA of multipath signals. From FIM, the error bounds for position and orientation were calculated. Their results show that the state of UE can be estimated with a sub-meter position error and a one-degree orientation error.
	
%	The idea of exploiting MPCs for localizing the mobile agent in indoor environments is reported in \cite{BEINSCHOB2017}. In this work, authors considered a fixed anchor node and a mobile agent, both equipped with multiple antennas. Measurements have been carried out at a carrier frequency of 63 GHz under LoS and Obstructed LoS (OLoS). The received signal by mobile agent contains transmitted signal, amplitude and delay of MPCs, Diffuse Multipath (DM) that causes interference to MPCs, and noise. The multipath-assisted localization algorithm called Simultaneous Localization and Mapping (SLAM) is used to map the environment containing the positions of VAs.  Moreover, a data association based on subpattern assignment is implemented to identify the signal sources (VA) and eliminate the false detections. The results show that the positioning error of 2 cm and 3 cm is achieved for LoS and OLoS situations.
	
	In general, the position of UE is estimated by assuming that both UE and the base station are time synchronized. But, this is not true all the time. In \cite{ABUSHABAN2020}, authors removed this assumption and estimated the position of UE using Two-Way Localization (TWL) model. Their model is based on two protocols: 1) Round Trip Localization Protocol (RLP), where the base station sends a signal to the UE, which, in turn, responds back to the base station with another signal after a pre-agreed time interval. Thereafter, the position and orientation of UE can be estimated. 2) Collaborative Localization Protocol (CLP), where the base station in addition to the response from the UE also receives another response through an error-free feedback channel. Based on the two received signals the localization is performed. The PEB and OEB for both protocols are calculated.  
	
	\subsection{5G Massive Multiple Input Multiple Output System}
	The massive Multiple Input Multiple Output (MIMO) concept uses multiple antennas equipped on both transmitter and receiver to detect and estimate the AoA of received signals accurately. The availability of temporal and spatial resolution in MIMO systems can be used to harness the position and orientation information of the UE from the NLoS components. In \cite{MENDRZIK2019}, authors used FIM to study the influence of NLoS on the UE state. %Authors derived the Equivalent FIM (EFIM) for position and orientation by integrating NLoS components.
	They showed that each NLoS component contributes to the positioning accuracy of UE, thereby, reducing the PEB and OEB. The results show that even in the absence of Line-of-Sight (LoS), UE position and orientation can be estimated with at least 3 NLoS components. Moreover, the beams from the base station should be narrow to illuminate the obstacles that generate MPCs.
	
	Authors in \cite{ZENG2021} proposed a new technique to estimate the speed, direction, and position of a moving UE. Their model is based on the time reversal focusing effect, where the power of the transmitted signal is focused at the intended location. Based on this focusing beam received by the UE, the Auto Correlation Function (ACF) is computed. Thereafter, the localization algorithm computes the speed, direction, and position of the UE by taking the ACF of the received signal and the geometry position of the base stations as inputs.
	
	\subsection{5G Ultra-Dense Network}
	The high-density deployment of base stations in a given industrial area allows UEs to have LoS with multiple base stations in most cases. Authors in \cite{MENTA2019}, proposed a UDN architecture-based edge cloud 5G system to estimate the UE positions. Their model uses AoA of LoS components to estimate the UE position through a two-stage approach. In the first stage, Extended Kalman Filter (EKF) at each base station estimates the AoA of the incoming signal from UE. In the second stage, EKF running in the edge cloud fuses the AoA estimates of all the base stations to estimate the UE position.
	%The authors demonstrated their model in a real scenario environment, where a large number of base stations connected to the lamp posts were deployed.
	The uplink signals sent from the UE to the base station are used to estimate the AoA. The Maximum Likelihood Estimator (MLE) and EKF are used to estimate the channel parameters. From the results, it is evident that EKF outperforms MLE, and the positioning accuracy of the sub-meter is achieved even at low frequencies by using UDN architecture.
	
	\subsection{5G Device-to-Device Communication}
	The D2D communication allows each UE in the network to talk to neighboring UEs using 5G communication links. A UE which is not connected to the base station can acquire the information from the adjacent UE. This enables cooperative positioning among UEs and improves accuracy and coverage area. In \cite{KIM2020}, authors proposed cooperative positioning and mapping method, where each vehicle in the environment uses Simultaneous Localization and Mapping (SLAM) to build the joint vehicle map density, which is further shared with the other vehicles through the base station, to enable cooperative positioning among vehicles. According to the authors, the source of the received signal can be a base station, reflecting surface, or a Scattering Point (SP). The authors assume that the location of the base station is known to the vehicle. The locations of the other two sources are found by locally running Probability Hypothesis Density (PHD) filter-based SLAM on each vehicle. At every instant of time, this filter predicts/updates the vehicle's trajectory and map information (locations of reflecting surfaces and SP). The information collected by each vehicle is sent to the base station in uplink transmission, where it fuses the information from all vehicles to create a global map of the environment. This global map is further shared with the vehicles in downlink transmission. 
	%Authors evaluated their model by deriving Mean Absolute Error (MAE) for vehicle trajectory, which showed promising results with sub-meter accuracy. 
	
	\subsection{Reconfigurable Intelligent Surface}
	At mmWave frequencies, the large available bandwidth provides high time resolution in delay estimate, because of which accurate ToA measurements for MPCs can be achieved. However, if the amplitude of MPC is too small, the ToA measurement of this MPC can lead to an error, due to which the performance of multipath-assisted localization degrades. To address this issue, Reconfigurable Intelligent Surface (RIS) assisted positioning has been proposed. RIS is a planar surface that consists of multiple small elements that can be programmed to control the radio waves impinging on it and directing them toward the UE by changing the phase of the reflected waves. In this way, RIS adds new controllable paths, thereby, improving positioning accuracy. Authors of \cite{LUO2022} developed the positioning model for RIS in near-field communications, where the large size RIS is decomposed into \textit{L} small segments with a size smaller than signal propagation distance. Therefore, the channels from the \textit{l}-th segment of RIS to the base station and the UE are in far-field communications. A new positioning framework called the coarse-to-fine algorithm was introduced, where the initial position estimate of UE is given by measuring the AoA of the incoming signal from the RIS segment. Followed by defining a rectangular area based on the coarse estimate and refining the positioning results through grid search.
	
	The overview of RIS-enabled localization for different scenarios was presented in \cite{KAMRAN2022}. In particular, the authors presented the RIS operation in downlink localization with respect to the number of base stations, wide-band signaling, and multi-antenna systems. Moreover, the comparisons between RIS and base station deployment scenarios were drawn in terms of cost, accuracy, and energy usage. According to the authors, RIS has low manufacturing costs compared to base stations as they can be manufactured with low-cost devices and can be mounted on buildings. Moreover, as RIS is envisioned as a passive beamformer, the energy consumption is low compared to that of a base station. In terms of accuracy, the large number of elements on RIS generates narrower beams compared to the base station with multiple antennas. However, the direct signal path from the base station to the UE has a low path loss than the one from a RIS. Therefore, other factors like operating frequency and network geometry need to be considered to draw a conclusion between the base station and RIS in terms of positioning accuracy. 
	
	Authors of \cite{ALEX2022} achieved sub-meter positioning accuracy of UE by deploying multiple RISs in place of base stations. In their model, each RIS consists of \textit{m} elements with a single receiver radio frequency chain for estimating the channel. A UE with a single antenna at an unknown position transmits the pilot sequences, which are received by each RIS. The baseband measurements collected at \textit{m}-th element of each RIS are sent to a controller, where the AoA of LoS component from each RIS is estimated using the Orthogonal Matching Pursuit (OMP) algorithm. From the AoA estimations, the UE position is extracted using MLE.

	\section{Shortcomings in recent works of 5G-based positioning}
	In this section, we present the shortcomings found in the recent works of 5G-based positioning and manifest the need to explore future research directions for the development of precise I4.0 indoor positioning technology. Our analysis was based on the AGVs point-of-view, which are deployed in challenging environments like industries.
	
	%\begin{comment}
	\FloatBarrier
	\begin{table*}[ht]
		\caption{Recent works on 5G-based positioning.}
		\centering
		\begin{tabular}{p{0.13\linewidth}p{0.09\linewidth}p{0.10\linewidth}p{0.10\linewidth}p{0.2\linewidth}p{0.10\linewidth}p{0.08\linewidth}}
			\hline
			Application & Environment & Frequency & Bandwidth & Method & Accuracy & Reference\\
			\hline
			UE positioning & Simulation & 60 GHz & 100 MHz & Coarse-to-fine & Millimeter & \cite{LUO2022}\\
			\hline
			Connected car & Simulation & 38 GHz & 125 MHz & Two-way localization & Sub-meter & \cite{ABUSHABAN2020}\\
			\hline
			UE positioning & Indoor & 60 GHz & Unknown & Line intersection & Sub-meter & \cite{KAMRAN2022}\\
			\hline
			UE positioning & Simulation & 28 GHz & Unknown & Auto correlation function & Centimeter & \cite{ZENG2021}\\
			\hline
			Connected car & Outdoor & 3.42 GHz & 15.36 MHz & Extended kalman filter & Sub-meter & \cite{MENTA2019}\\
			\hline
			UE positioning & Simulation & 30 GHz & Unknown & Maximum likelihood estimator & Sub-meter & \cite{ALEX2022}\\
			\hline
			Connected car & Simulation & mmWave & Unknown & PHD filter & Sub-meter & \cite{KIM2020}\\
			\hline
			Connected car & Simulation & mmWave & Unknown & EK-PHD filter & Sub-meter & \cite{KALTIOKALLIO2021}\\
			\hline
			Connected car & Simulation & mmWave & Unknown & PHD-SLAM & Sub-meter & \cite{KIM2022}\\
			\hline
			UE positioning & Simulation & 30 GHz & Unknown & Carrier phase positioning & Centimeter & \cite{FAN2022}\\
			\hline
			UE positioning & Simulation & 28 GHz & 20 MHz & Geometric channel & Sub-meter & \cite{WEN2021}\\
			\hline
			UE positioning & Indoor & 220 GHz & Unknown & SLAM & Centimeter & \cite{ALADSANI2019}\\
			\hline
			UE positioning & Simulation & mmWave & Unknown & Message passing & Sub-meter & \cite{RICO2019}\\
			\hline
		\end{tabular}
		\label{tab2}
	\end{table*}
	%\end{comment}
	
	\subsection{Simulation environment}
	The majority of research works on 5G-based positioning assessed the performance of their methods and models in a simulated environment \cite{KALTIOKALLIO2021, KIM2022}. The challenge of using these simulation environments is that they cannot capture the real world perfectly, leading to unrealistic conditions and the results obtained cannot covenant the same experience in the real world. Some of the factors that have been ignored in the previous works based on simulations are: 1) the path-loss model does not consider the attenuation due to walls and other objects in the surrounding environment, 
	%2) the simulated area used for demonstrating the efficiency of the proposed model is limited in size,
	2) the positioning of UE was done by considering only the base stations that are in LoS with the UE, 3) the number of objects (reflectors and scatters) considered in the environment is limited, 4) the positioning algorithm considers only the specular multipath while ignoring the Diffuse Multipath (DM), 5) the UE and base stations are assumed to be perfectly synchronized, 6) the network deployment model for deploying the base stations was not considered  \cite{FAN2022, WEN2021}. In contrast to the unrealistic assumptions made in simulations, the industrial environment consists of multiple base stations deployed over a large area having objects of different shapes and roughness values, giving rise to specular and diffuse reflections, which reach the AGV in more than one path. In the literature, there is a lack of investigation of the 5G-based positioning accuracy in industrial environments.
	
	\subsection{Simultaneous Localization and Mapping}
	The availability of high bandwidths and large antenna arrays in 5G allows for the resolution of the MPCs using time and angular domains. Furthermore, these MPCs can be mapped to the physical objects present in the environment, creating a digital map. Such maps can be used to localize the AGV through a method called Channel-SLAM \cite{GENTNER2016}. However, there are certain challenges involved in using Channel-SLAM. First, each AGV has a certain Field-of-View (FoV), up to which it can detect the objects. There is a probability that even the objects that are inside the AGV FoV can go undetected due to the receiver's imperfections. Secondly, the misdetection of objects due to the effect of noise on channel estimation. Lastly, the data-association problem corresponds to identifying the origin of MPCs. 
	%Lastly, the map generated by each AGV has to be shared with neighboring AGVs to enable cooperative positioning. 
	To address these challenges of SLAM, different methods like geometry-based, message passing, and Random Finite Set (RFS) have been proposed in the literature. In \cite{ALADSANI2019} and \cite{RICO2019}, geometry-based and message-passing methods are used for the SLAM. However, the problem of data association is not considered by the authors in these works. In \cite{KIM2020}, authors addressed the above challenges of SLAM using RFS theory. In their model, authors employed the PHD filter to map the objects in the environment and classified them as the base station, SP, and Virtual Anchor (VA). 
	%The results obtained from each vehicle are fused at the base station and the result of the updated map is sent to the vehicle fleet in the downlink.
	However, the authors evaluated their model using simulation and considered only a single base station. The efficiency of their model in large-scale scenarios like industries was not investigated.           
	
	\subsection{Network infrastructure geometry}
	The general way of estimating AGV position with radio signals is by using range-based methods, where signals from the base stations located at known positions are used by the positioning algorithm. In previous cellular networks, where the distance between base stations and UE is in the order of a few kilometers, the positioning accuracy used to be affected mainly by the Signal-to-Noise Ratio (SNR) of the received signal at UE. On the other hand, in 5G networks with UDN, it is now possible to have a much shorter distance between the base station and UE, a higher probability of LoS, and a high density of base stations. In such a scenario, the only factor that influences the positioning accuracy will be the network geometry. The sub-meter positioning accuracy cannot be achieved by using 5G UDN alone but through an effective deployment model of base stations, where at least three base stations will be involved in positioning the UE \cite{ROTH2019}. The 5G network should be designed and deployed by keeping the above points in mind. Some of the common infrastructure deployment models used in literature are Poisson Point Process (PPP) and Hard Core Point Process (HCPP). However, the research works on 5G-based positioning haven't considered the deployment models for the UDN, despite its correlation to positioning accuracy. %Furthermore, the perfect network deployment model for a 5G UDN is still a open question. 
	
	\section{Enhanced Techniques}
	The 5G technology alone cannot satisfy the accuracy requirements of AGV in the industrial environment. Therefore, it is necessary to work on new possibilities for improving positioning accuracy by complementing the 5G technology with other technologies. In this section, we present enhanced techniques that can enhance the 5G-based positioning accuracy, while remaining in congruence with the 5G system specifications. Fig.~\ref{figure1} shows the proposed positioning architecture for AGV.
	
	%\begin{figure}[t]
		%\centering
		%\includegraphics[scale=0.47]{System_model.PNG}
	%	\includegraphics[clip, trim=2cm 5cm 0.5cm 5cm, width=0.53\textwidth]{system model.pdf}
	%	\caption{Proposed positioning architecture for AGV.}
	%	\label{figure1}
	%\end{figure}
	%Several of the use cases presented by industrial associations refer to extremely stringent latency and accuracy requirements, which might not be met by the 5G technology alone, especially in challenging operating conditions. In these scenarios, it is therefore necessary to employ advanced localization and sensing techniques, and to hybridize with non-radio technologies, while remaining in accordance with the 5G architecture
	\subsection{Sensor Fusion}
	%In the process of estimating an unknown parameter, measurements from different sensors can be fused together to enhance the estimation accuracy. This is known as sensor fusion.
	The accuracy of 5G-based positioning can be improved by combining the measurements with the exteroceptive and proprioceptive sensors. A proprioceptive sensor like the Inertial Measurement Unit (IMU) is used to track the positions of AGV at regular intervals. On the other hand, exteroceptive sensors like Light Detection and Ranging (LIDAR) and camera are used to sense the environment around AGV through SLAM by creating a map of obstacle positions to avoid collisions \cite{ALVARADO2018}. By combining multi-modal sensors data with 5G measurements data the positioning error can be reduced because these sensors deployed on the AGV observe the same environment and are used to sense different parameters (acceleration, angular velocity, distances). These parameters hold the location information and when processed correctly can beneficially contribute to the position estimation. However, when the sensor data need to be fused with the 5G measurements data, an integration problem arises. The loosely coupled fusion results in independent measurements and provides two different position estimations. To overcome this problem, efficient data fusion algorithms are needed.
	
	\subsection{Context Information}
	The context information available in the form of industrial processes can be used to improve the position prediction of the current observer, AGV. This additional information is provided to AGV before it begins its operations in the form of a map and can also be obtained by AGV itself during its dynamic motion. In the former case, the locations in the industry consisting of obstacles (objects, other UEs) with distinctive features and also areas that result in the unique pattern of received signals (RSS, ToA) due to the presence of heavy metallic beam-like structures can act as prior knowledge. These locations can be used as landmarks to correct the 5G positioning errors. The landmarks should be static and also be available at all times for the AGV while it is moving. But, it does not need to be a real physical structure. The VA, formed from the mirror image of the real anchor can also be used as a landmark because these VAs are time-invariant and are synchronized with real anchors. Each MPC can be assumed to be transmitted from VA, in LoS scenario. Therefore, when AGV is moving around the industry it can make use of VAs to improve its position prediction. 
	%However, efficient data-association algorithms that can map MPCs to the correct VAs are needed. %The signal transmitted from the anchor bounces through different obstacles and reaches AGV via different paths, known to be MPCs.
	
	\begin{figure}[t]
		\centering
		\includegraphics[clip, trim=2cm 5cm 0.5cm 5cm, width=0.41\textwidth]{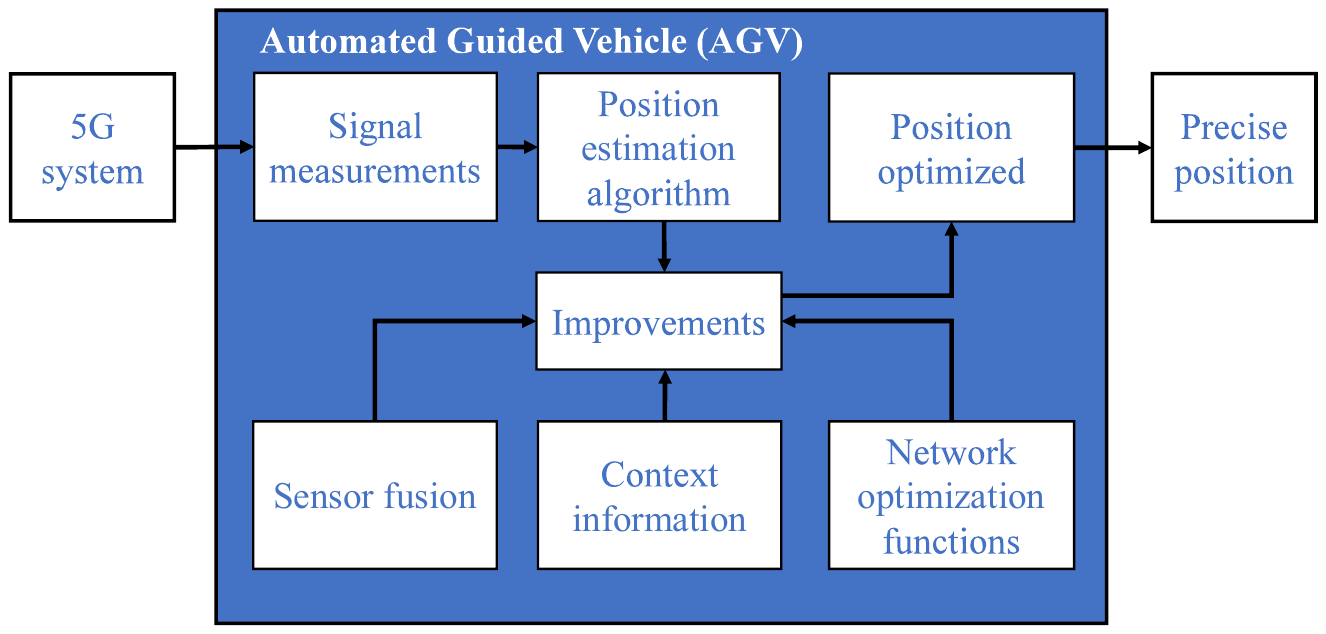}
		\caption{Proposed positioning architecture for AGV.}
		\label{figure1}
	\end{figure}
	
	\subsection{Network Optimization}
	Efficient network optimization functions are necessary to achieve precise positioning accuracy in 5G networks. The radio resources (such as transmitting power, transmitting time, and signal bandwidth) allocated to base stations will have a great impact on positioning accuracy. For instance, two AGVs exchanging ranging information with the base station simultaneously can cause packet collision. Moreover, AGV communicating with the base station using poor link quality can result in measurement errors. To address these issues, network functions like node activation, node prioritization, and node deployment need to be employed. The goal of the node activation function is to select a certain number of AGVs in the network that are allowed to make measurements with the base stations at a particular time instant in order to avoid packet collisions and minimize the positioning error. The goal of the node prioritization function is to ensure that each AGV in the network is provided with sufficient radio resources to make measurements with the base stations as well as with the neighboring AGVs. The goal of the node deployment function is to determine the positions of new base stations in the network such that the performance and positioning accuracy of existing base stations can be maximized. Moreover, network topology optimization by seamless activation and deactivation of certain base stations in a UDN can positively impact the accuracy of the measurement data by reducing interference effects \cite{TORRIERI1984}.  
	
	\subsection{Reconfigurable Intelligent Surface Assisted Positioning}
	The NLoS condition between the transmitter and receiver is a very common problem to occur in large-scale environments like industries. Though the MPCs resulting from NLoS condition can be exploited by the 5G system for positioning, it all depends on the quality of MPCs. RIS has evolved as a promising technology for controlling signal propagation and customizing the environment to 5G SLAM requirements. Compared to the traditional wireless repeaters used in cellular systems, which can also control signal propagation, the RIS comes with low hardware complexity and cost as it does not require amplifiers \cite{BJORNSON2020}. To this end, we believe that the deployment of RIS in industries can further enhance the 5G SLAM accuracy. However, certain challenges related to RIS-assisted positioning need to be addressed. First, the channel model with multiple scatterers for the indoor environment needs to be modeled. Second, to direct the reflected beam toward the AGV, each element of RIS has to be configured with a particular phase shift value. However, identifying the optimal phase shift value from a large number of possible values (between $+$90\textdegree \hspace{1mm}and $-$90\textdegree) is challenging \cite{BJORNSON2020}.
	
	\section{Conclusion and Future Work}
	This paper surveys the recent advances in 5G-based positioning with a particular focus on its enabling features. For each enabling feature, we provide a review of methods recently proposed, along with their shortcomings. From our analysis, the current solutions lack an evaluation of their performance level on a large-scale and in live production environments. Finally, we propose some enhanced techniques that can complement the accuracy of 5G-based positioning for the use case of AGV. In our future work, we will focus on these techniques to concept improved positioning technologies for I4.0.
	
	\section*{Acknowledgment}
	This work has received funding from the European Union's Horizon 2020 research and innovation programme under the Marie Sklodowska-Curie grant agreement ID 956670. Also, partially funded through the project ROUTE56 (funded by the Agencia Estatal de Investigación, PID2019-104945GB-I00~/~AEI~/~10.13039/501100011033).

	\bibliographystyle{IEEEtran}
	\bibliography{IEEEabrv,references}
	
\end{document}